\documentclass{svmult}

\usepackage{makeidx}         % allows index generation
\usepackage{graphicx}        % standard LaTeX graphics tool
\usepackage{multicol}        % used for the two-column index
\usepackage[bottom]{footmisc}% places footnotes at page bottom
\makeindex             % used for the subject index
\begin{document}

\title*{RHESSI Results -- Time For a Rethink?}
\author{J. C. Brown\inst{1}
\and E. P. Kontar\inst{1} \and A. M. Veronig\inst{2}}
\authorrunning{J. C. Brown et al.}
\institute{Department of Physics and Astronomy,
Kelvin Building,
University of Glasgow,
Glasgow G12 8QQ,
Scotland, UK
\texttt{john@astro.gla.ac.uk,eduard@astro.gla.ac.uk}
\and
Institute for Geophysics, Astrophysics and Meteorology,
University of Graz,
Universit\"atsplatz 5,
A-8010 Graz,
Austria
\texttt{asv@igam.uni-graz.at}
}
\maketitle

\abstract{ Hard X-rays and $\gamma$-rays are the most direct
signatures of energetic electrons and ions in the sun's atmosphere
which is optically thin at these energies and their radiation
involves no coherent processes. Being collisional they are
complementary to gyro-radiation in probing atmospheric density as
opposed to magnetic field and the electrons are primarily 10--100~keV 
in energy, complementing the ($>$100~keV) electrons likely
responsible for microwave bursts.

The pioneering results of the Ramaty High Energy Solar
Spectroscopic Imager (RHESSI) are raising the first new major
questions concerning solar energetic particles in many years.
Some highlights of these results are discussed -- primarily around
RHESSI topics on which the authors have had direct research
involvement -- particularly when they are raising the need for
re-thinking of entrenched ideas. Results and issues are broadly
divided into discoveries in the spatial, temporal and spectral
domains, with the main emphasis on flare hard X-rays/fast
electrons but touching also on $\gamma$-rays/ions, non-flare
emissions, and the relationship to radio bursts. }


\section{Introduction}
\label{intro} Major observational results from RHESSI and
instrumental details have been extensively described elsewhere (e.g.
\cite{lin:al-02} and other articles in that volume, and
\cite{den:al-05}) and will not be repeated here.  Based on results from
numerous earlier spacecraft from OGOs, OSOs and TD1A through SMM,
Hinotori and Yohkoh (these three giving the first HXR images), the
conventional wisdom prior to RHESSI envisaged electron and ion
acceleration high in a loop near a reconnection site.  Most of the
hard X-rays (HXRs) and $\gamma$-rays were believed to originate in
two bright loop footpoints by collisional thick target deceleration
of fast particles with a near power-law spectrum in the dense
chromosphere \cite{bro-71}, plus occasional fainter emission at or
above the looptop as seen in Yohkoh \cite{mas:al-94} and sometimes
even higher as seen in limb occulted flares \cite{kan-83}.  Until
RHESSI, apart from one balloon flight \cite{lin:sch-87}, spectral
resolution was very limited, particularly in images and in
(non-imaged) $\gamma$-rays.  RHESSI has transformed this via Ge
detector spectrometry, yielding high resolution spectra and spectral
images in HXRs, high resolution $\gamma$-ray line spectroscopy, and
the first $\gamma$-ray line images.  RHESSI also excels in having an
unsaturated spectral range from a few keV to ten of MeV, thus
yielding data on the hot SXR plasma as well as on fast particles
(see articles in special issues of Solar Phys.\ vol.\ 210, 2002 and
Astrophysical Journal Letters vol. 595, 2003). While many of the
RHESSI data show events with some resemblance to the canonical thick
target footpoint scenario, with near power-law spectra, there are
many examples deviating from this simple picture.  Here the main
emphasis is on these new features as they are the driving force
behind the need for a rethink.

\section{Imaging Discoveries and Issues}
\label{image}

\begin{figure}[tb]
\centering
\includegraphics[height=7cm]{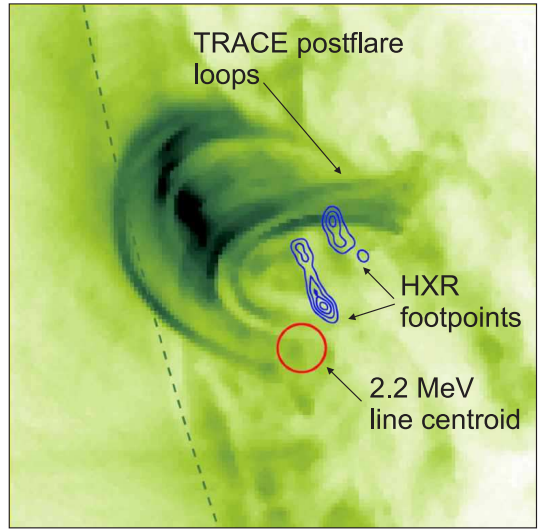}
\caption{Hard X-ray (electron) emission versus 2.2 MeV (nuclear)
gamma-ray line emission centroid location for July 23, 2002 with
TRACE context, based on \cite{hur:al-03}. The displacement of the fast
ions from the fast electrons was one of the biggest surprises in
RHESSI data.
}
\label{hurford_gammas}
\end{figure}

Probably the most exciting imaging discovery by RHESSI is the fact
that, in at least one of the few strong $\gamma$-ray line events
seen by RHESSI, the 2.2 MeV neutron capture line comes from a
spatial location quite distinct from the source of HXRs --
Figure~\ref{hurford_gammas} \cite{hur:al-03}. Cross-field transport
is unable to explain this spatial separation and it seems it must be
due to acceleration of electrons and of ions in or into quite
distinct magnetic loops.  The only explanation offered to date is
that by Emslie, Miller and Brown \cite{ems:al-04} where the
longer/shorter Alfv\'{e}n travel time in larger/smaller loops
favours respectively the stochastic acceleration of ions/electrons.
The Doppler profiles of the $\gamma$-ray lines also help constrain
the geometry of the loop in which the same ions move.

\begin{figure}[tb]
\centering
\includegraphics[height=7.5cm]{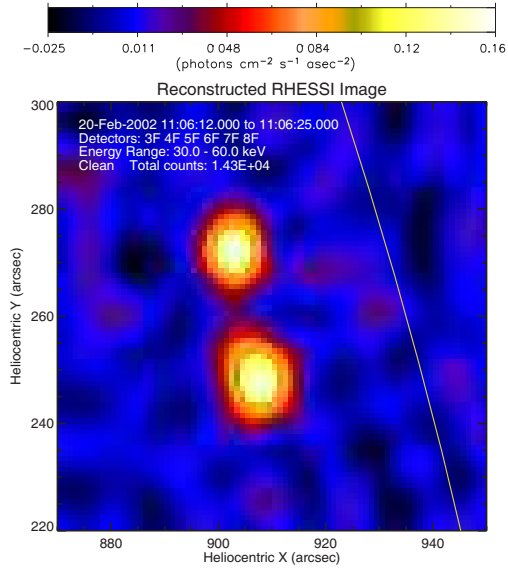}
\caption{Simple `classic' 2-footpoint flare seen by RHESSI. While there
may be other interpretations, such a structure is expected if the field
approximations to a simple bipolar loop without strong field
convergence. Then electrons accelerated anywhere in the upper, 
low-density, loop can reach the dense chromosphere at the loop ends 
(`footpoints') where they emit bremsstrahlung very strongly compared 
with in the tenuous corona. Such high `footpoint contrast' was discussed 
as early as Brown and McClymont \cite{bro:mcc-76} and MacKinnon, Brown 
and Hayward \cite{mac:al-85}.
}
\label{xrayfeet}
\end{figure}

\begin{figure}[tb]
\centering
\includegraphics[height=8cm,clip]{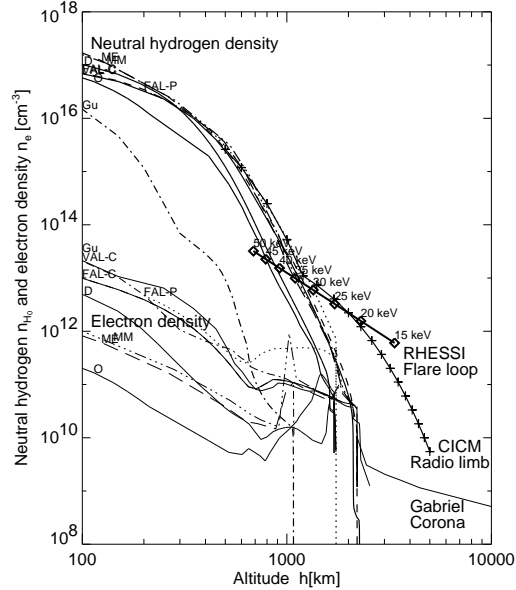}
\caption{Height distribution of hydrogen and other densities, as
labelled, in numerous solar atmospheric models with superposed that
required for a collisional thick target to match the RHESSI data for
the flare of 2002, February 20  from \cite{asc:al-02}. The basis of
\cite{asc:al-02} is to assume that `footpoint flares'
like that in Figure 2 confirm the collisional thick target model of
injection of electrons from the corona down the legs of a loop where
they undergo purely collisional transport as they radiate. Since high
energy electrons penetrate deeper, the footpoint centroid height should
decrease with increasing energy and at a rate depending on the plasma
density there. For an assumed electron injection spectrum one can 
then use the energy dependence of the HXR centroid height to infer 
the density as a function of height, the beam acting as a probe of the 
target.
}
\label{denstruct}
\end{figure}

Those events which do show a classic 2-footpoint structure (e.g.
Figure~\ref{xrayfeet}), at least within the spatial resolution
limits of RHESSI, can in principle be compared with the predictions
of the thick target model in terms of the spectral variation of the
footpoint structure. Qualitatively, the highest energy electrons
penetrate deepest so that the hardest HXR footpoints should lie
lowest, and furthest apart in the loop, to an extent depending on
the variation of electron energy loss rate with electron energy.  On
the conventional assumption of collision-dominated energy losses,
Brown, Aschwanden and Kontar \cite{bro:al-02} and Aschwanden, Brown
and Kontar \cite{asc:al-02}, determined the atmospheric density
structure $n(h)$ needed for the thick target model to produce the
observed spectral image structure, with the results shown for one
event in Figure~\ref{denstruct} (February 20, 2002; for a HXR 
map see Figure~\ref{xrayfeet}). These show that collisions are a
substantial factor in electron transport and may be
the dominant factor if flare $n(h)$ is similar to spicules
(Figure~\ref{denstruct}).  A lower $n(h)$ structure requires some
non-collisional energy losses to fit the model to the data. This
modeling needs improving to allow for pitch angle changes and for
the variation of collision cross section as the target ionisation
decreases, before firm conclusions are drawn.

\begin{figure}[tb]
\centering
\includegraphics[height=10cm]{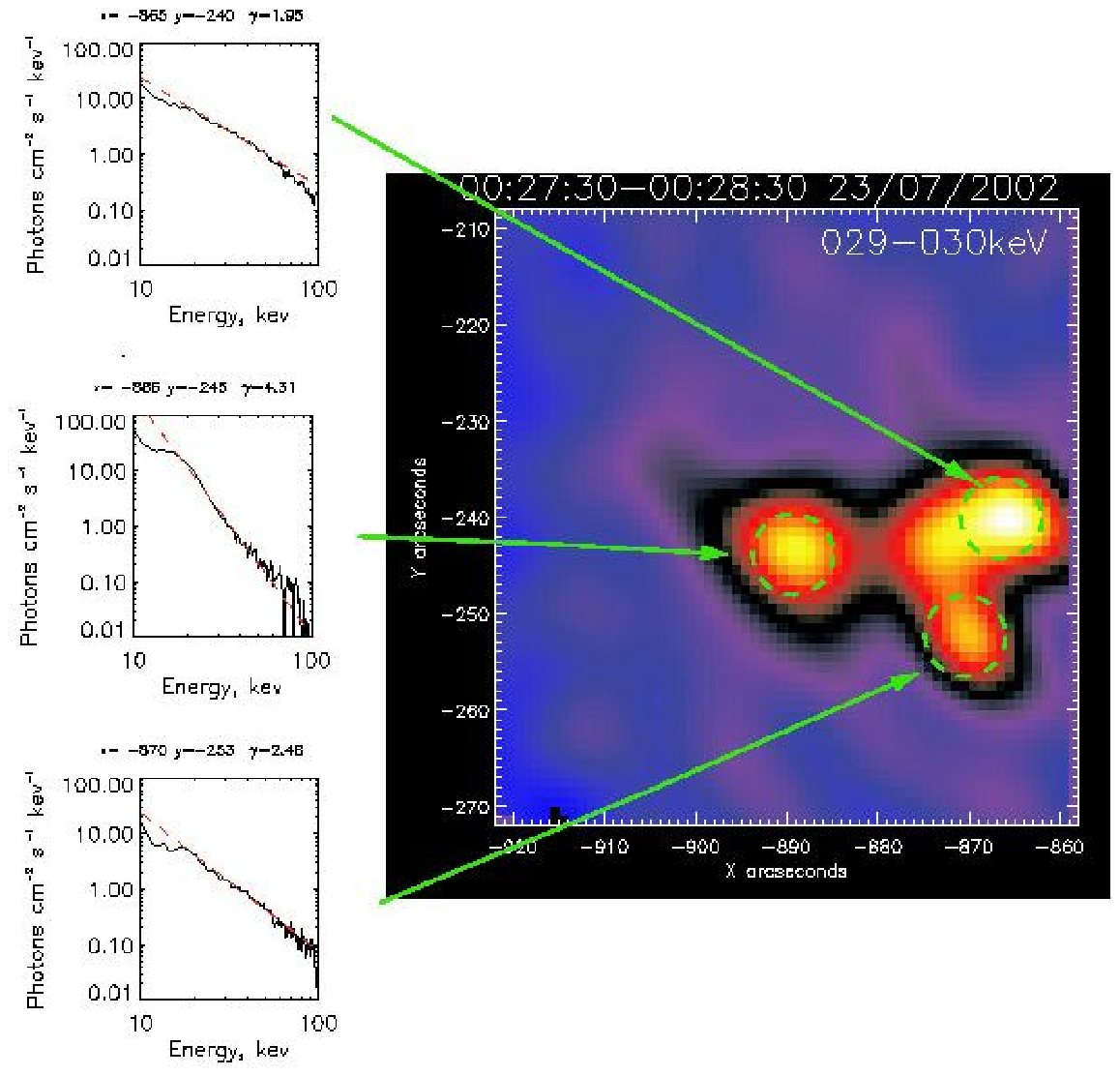}
\caption{Complex HXR spectral image of the 2002 July 23 flare from
\cite{ems:al-03}. This event seems to show four distinct sources and
does not conform to the simple bipolar pattern of Figure 2. If, for 
example, the two rightmost patches were footpoints of a single loop and 
the leftmost one the looptop, then their
local spectral indices -- measurable accurately for the first time by
RHESSI spectrometric imaging -- are inter-related roughly as expected in
the collisional thick target model, the footpoint spectra being roughly
2 powers harder than the looptop. But the looptop source is higher than
expected and the fourth source is hard to explain in any simple way --
cf. \cite{ems:al-03}. RHESSI's spectral resolution, sensitivity,
and large dynamic range are enabling such questions to be asked for the
first time. The evolution of the different RHESSI sources
superposed on an MDI magnetogram is shown in Figure~11 of the article by 
Dennis et al. \cite{den:al-05} in this volume.
}
\label{23julemslie}
\end{figure}

Some events show more complex HXR structure though this is in part due
to higher photon fluxes enabling detection of fainter components.  An
example is the extensively studied event of July 23, 2002, already shown
in broad context in Figure~\ref{hurford_gammas}. Figure~\ref{23julemslie}
shows that in the deka-keV HXR
range the source comprises two bright footpoints, with hard spectra, and
possibly a third, or at least one extended, footpoint, with a distinct
fainter and softer source, possibly at or near the looptop.  This event
is one of those subjected most thoroughly so far to spectral image
reconstruction \cite{ems:al-03} though that facet of RHESSI data
reduction is still being refined.

\begin{figure}[tb]
\centering
\includegraphics[height=7cm]{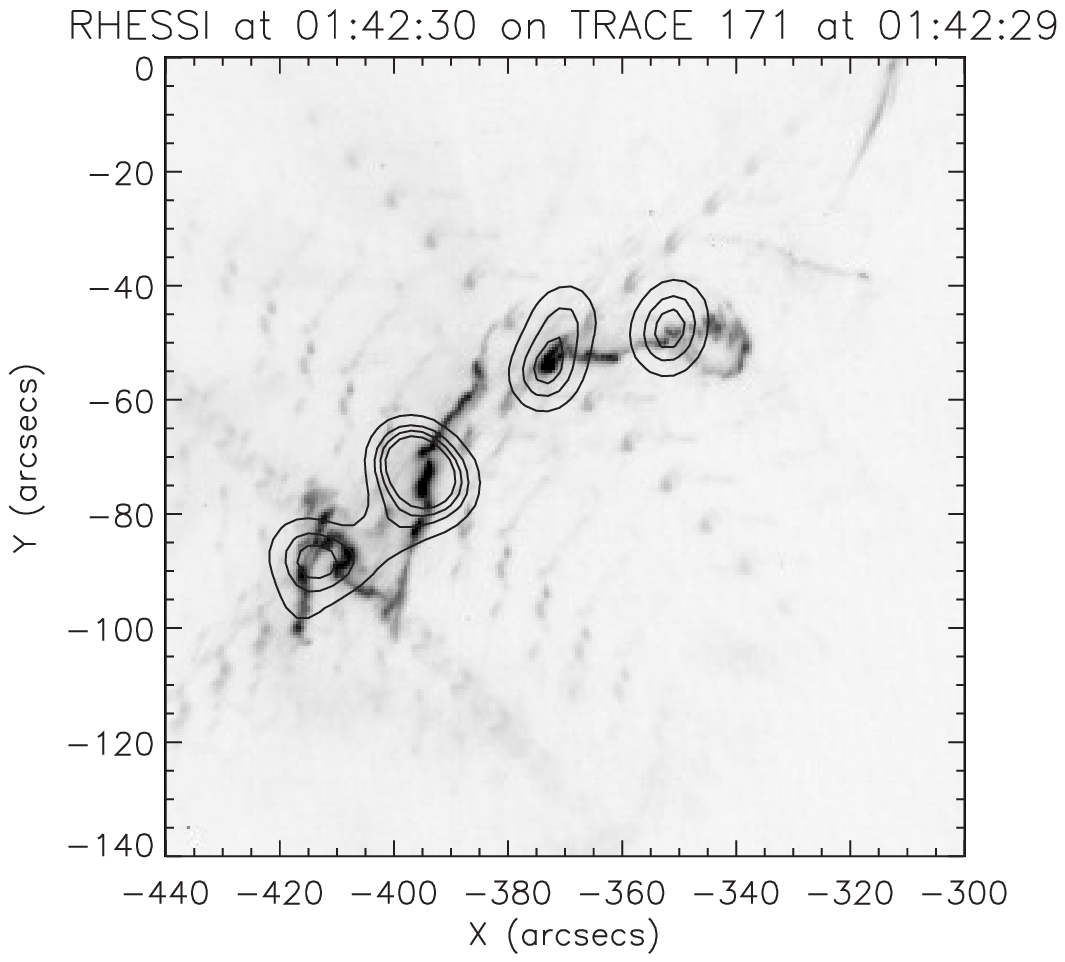}
\caption{RHESSI 30--50 keV contours overlayed on a TRACE 171~{\AA}
image of the peak of the M5.4 flare of 14 March 2002 from
\cite{fle:hud-02}. The RHESSI images were reconstructed with CLEAN using 
grids 3 to 9 giving an angular resolution of $\sim$8$''$. At the available 
spatial resolution, there is a good correspondence between the HXR sources 
and the TRACE 171~{\AA} kernels though the HXR patches are rather large,
and at any moment occupy only a small part of the overall flare ribbon
extent
observed in TRACE.
}
\label{fletcher_footpoints0}
\end{figure}

\begin{figure}[tb]
\centering
\includegraphics[height=6cm]{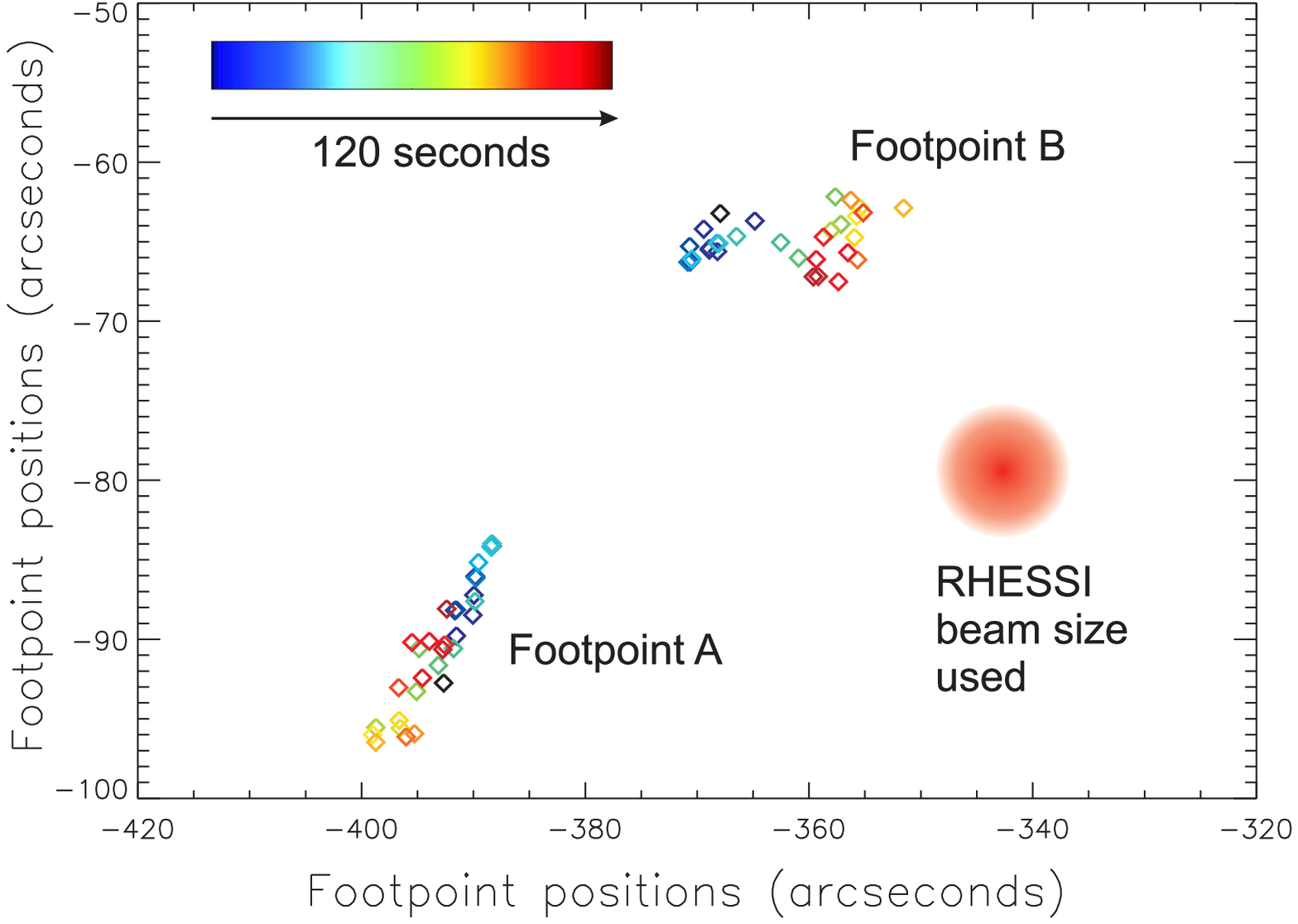}
\caption{Hard X-ray footpoint centroid motion in the M5.4 flare of
14 March 2002 from \cite{fle:hud-02}. Footpoint locations derived
from RHESSI 30--50~keV images are color-coded in time.
Comparison with TRACE XUV images (cf. \cite{fle:hud-02};
see also Figure~\ref{fletcher_footpoints0}) reveal that the HXR source
motions are perpendicular to as well as along the XUV flare ribbons
indicating that the HXR footpoint progression is much more complex than 
expected from simple 2D reconnection models.
}
\label{fletcher_footpoints}
\end{figure}

Fletcher and Hudson \cite{fle:hud-02} have studied the location and
motion of RHESSI HXR `footpoints' and compared them with those seen
by TRACE in the XUV range and at other wavelengths
(Figures~\ref{fletcher_footpoints0} and \ref{fletcher_footpoints}).
This reveals a relatively complex situation, as yet to be properly 
understood. The HXR patches are rather large, being seen as extended 
even at this limited resolution ($\sim$8$''$) but nevertheless at any 
moment occupied only a small part of the overall flare ribbon extent
(Figure~\ref{fletcher_footpoints0}), their centroids moving along it 
as the flare progressed (Figure~\ref{fletcher_footpoints}). The 
TRACE brightpoints are much smaller/better resolved and some of them 
more or less track the HXR patch centroids.  Fletcher and Hudson 
interpret this motion as reflecting the progression of the momentarily 
reconnecting field lines which direct particles to the chromosphere. 
While this may be true, and their observed sizes the result of `motion' 
of even smaller sources during the integration time, it is impossible 
in the conventional thick target model for HXR source electrons to be 
concentrated in regions anywhere near as small as a bundle of field 
lines near the very thin reconnection sheet. An intense burst
requires a beam rate $n_bv_bA\approx 10^{36}$\,s$^{-1}$ or a beam density
$n_b\approx 10^{11}/A_{15}$ where $A=10^{15}A_{15}$\,cm$^2$ is the beam area 
which requires an impossible large $n_b$ for $A \ll 10^{15}$ 
($\sim$1$''$ square).

\begin{figure}[tb]
\centering
\includegraphics[height=6.4cm]{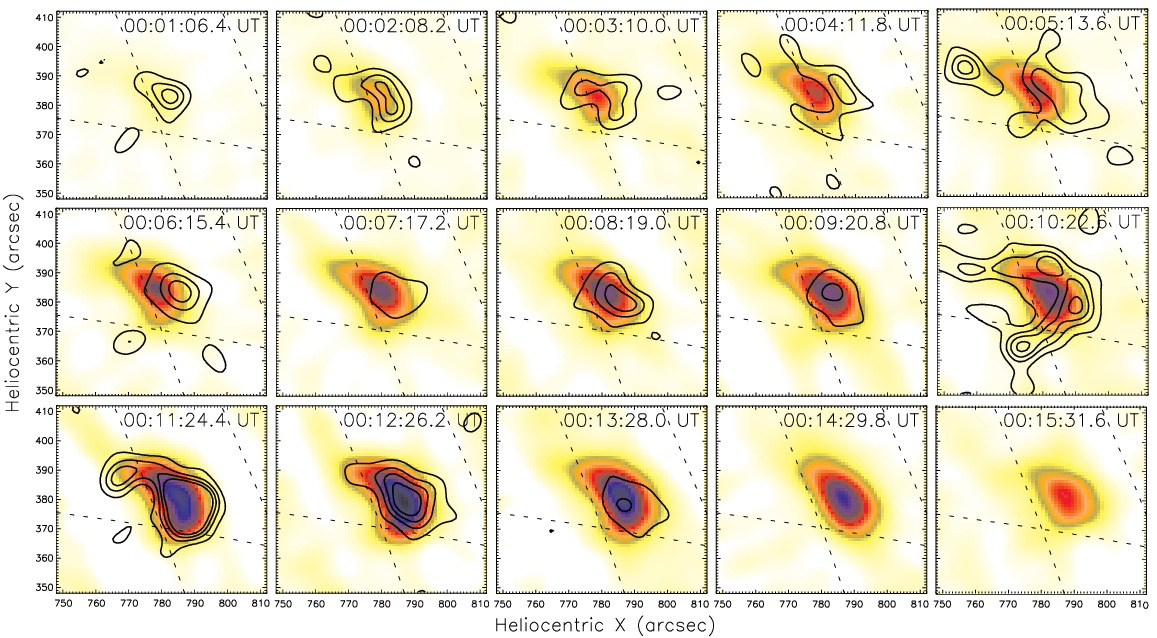}
\caption{RHESSI image sequence for the M3.2 flare of 2002 April
14/15, from \cite{ver:bro-04}. The  images show 6--12~keV, the contours 
25--50~keV RHESSI maps reconstructed with CLEAN using grids 3 to 8 (except 7). 
The soft 6--12 as well as the hard 25--50 keV emission are concentrated near 
the loop top. Only during the impulsive rise (00:01--00:06 UT) and 
briefly during the late highest peak (00:10:22 UT), is weak
footpoint emission detectable in the 25--50~keV band. Note that
throughout the event the loop top is the predominant HXR source, whereas in 
`normal' events footpoint emission prevails at high energies.
}
\label{veronig}
\end{figure}

Veronig and Brown \cite{ver:bro-04} discovered a new class of bright
coronal HXR source in which the loop (top) emission is hard and
dominates to high energies, with little or no emission from the
footpoints, in sharp contrast to Masuda coronal sources where the
footpoints still dominate -- Figure~\ref{veronig}. Similar RHESSI
events have been also studied by Sui et al. \cite{sui:al-04}.

Veronig and Brown interpret this as due to a high loop density
($>$\,$10^{17}$~m$^{-3}$), consistent with the soft X-ray emission
measure estimate $(EM/V)^{1/2}$, the coronal loop being then a
collisionally thick target at electron energies up to 50--60 keV.
The high loop density is also consistent with conductive evaporation
driven by collisional heating in the loop top.  Such an
interpretation had in fact previously been hinted at for Yohkoh data
\cite{kos:al-94} but such events seem to be rare and more easily
seen with RHESSI's large dynamic and spectral range.

An even more extreme class of high altitude source has been reported
by Kane and Hurford \cite{kan:hur-03} where there is an elevated, long 
lasting, HXR source seemingly `detached' from chromospheric emission. Kane,
McTiernan and Hurley \cite{kan:al-04} report a particularly
interesting case of coronal HXR emission seen by RHESSI for an
occulted flare behind the limb, but wholly seen by Ulysses which was
behind the sun.

The coronal source yields a significant fraction of the total flare HXR
flux showing the presence of copious fast electrons somehow confined
high in the atmosphere, somewhat akin to the long duration high altitude
$\gamma$-ray events studied earlier by Ramaty et al. in \cite{ram:al-97}. 
Much work remains to be done on the detailed quantitative modelling and 
physical interpretation of this class of HXR event, using the much more
comprehensive data available from RHESSI, Nobeyama etc. than was
possible in Kane's earlier ground-breaking stereo event studies
\cite{kan-83}.

\section{Temporal Domain Discoveries and Issues}
\label{timing}

The most important temporal information in RHESSI data is bound to
be in the evolution of the spatial and spectral characteristics, as
opposed to global light curves in single energy bands.
Unravelling the raw data to produce an X-ray `multi-colour' movie at
high spectral and spatial resolution is computationally very
demanding and can only be even remotely contemplated for intense
events with ample photons. However, a great deal can be gleaned from
more rudimentary temporal information such as comparison of image
sequences in two well separated energy bands, whole sun light curves
as a function of energy over hitherto unexplored energy ranges, and
comparison of light curves/image sequences at soft and hard X-ray
energies with data at wholly different energies.  In these
categories, among the `rethink' provoking RHESSI discoveries are the
following.

\begin{figure}[tb]
\centering
\includegraphics[height=8.5cm]{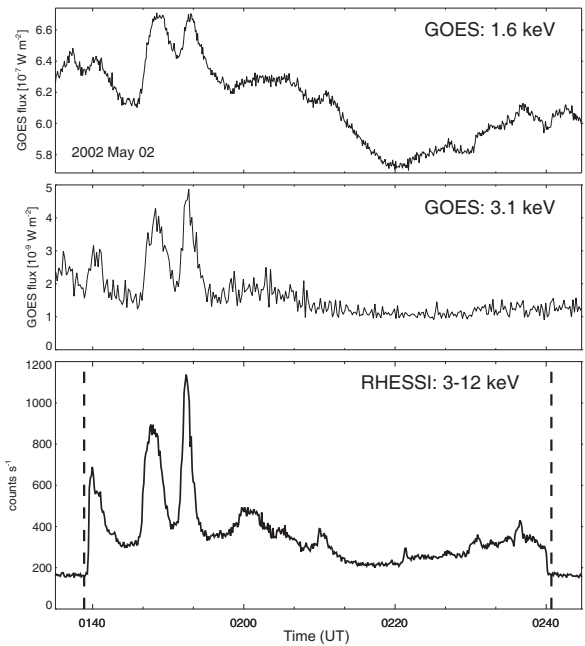}
\caption{Low energy HXR micro-events reported by \cite{kru:al-02}
showing an early RHESSI discovery that even the non-flaring sun
undergoes impulsive low energy but spectrally hard `HXR' events
quasi-continuously (intervals $\sim$ minutes).
}
\label{microevents}
\end{figure}

There has long been interest in the possibility of `nano'- or
`micro'- flares being an ongoing solar coronal phenomenon, possibly
involved in coronal heating.  Much research in this area has been
statistical in character but some papers have addressed the physics
of micro-events including their possible role in supplying mass to
the corona, as well as heating it.  Brown et al. \cite{bro:al-00}
claimed that micro-events in loops were not hot enough to provide
their emission measure increase by conductive evaporation of the
chromosphere.  They proposed that energetic electrons of around 10
keV might instead be responsible and predicted that RHESSI might
detect frequent low energy `hard' X-ray micro-events from the
non-flaring sun.  One of RHESSI's early discoveries was indeed that
the `non-flaring' sun exhibits micro HXR events of minutes' duration
at intervals of several minutes -- Figure~\ref{microevents}
\cite{kru:al-02}.  The detailed physics of these remains to be
investigated quantitatively but Krucker's movies ({\tt \small
http://sprg.ssl.berkeley.edu/\~{}krucker}) show them to be related
to XUV surges seen by TRACE, and to Type III radio bursts and
associated plasma waves, originating near the boundaries of HXR
sources.

The interplay of spectral, spatial and temporal information from
RHESSI is particularly clearly shown in the study by Veronig et al.\
\cite{ver:al-05} of the Neupert effect.  The empirical Neupert
effect is that the SXR light curve of a flare is well correlated
with the time integral of the HXR light curve \cite{neu-68, den:zar-93}.
The canonical interpretation is that the SXR plasma is heated by
cumulative energy input from HXR emitting fast electrons. The Neupert 
effect has been observationally established for the impulsive phase of many
flares, but in general the correlations are far from being perfect
\cite{den:zar-93,lee:al-95,mct:al-99,ver:al-02}.
These deviations were interpreted as the effect of plasma cooling
and/or grounded on the idea that the {\it physical} Neupert effect
should exist between the nonthermal and thermal energies and not between
the HXR and SXR lightcurves.
Veronig et al.\ quantitatively addressed this issue and suggested
that the imperfect temporal correlation may be due to the fact that
SXR flux depends on density and temperature, and not just on energy
content, that HXR flux depends on beam spectrum as well as power,
and that one must take account of plasma cooling by radiation and
conduction during the event.  There has of course been
considerable theoretical/numerical work \cite{li:al-93, li:al-97} to
see whether and how a Neupert effect is seen when one runs a model of
the impulsive heating of a loop and follows its evolution allowing for
hydrodynamics, evaporation, radiative and conductive cooling etc. The
broad answer is, unsurprisingly, that one does, the Neupert effect being 
clearest when the loop takes longest to cool. However, such theoretical 
models are {\it not} the same thing as testing to see if the Neupert 
effect {\it in real data} can be physically attributed
solely to beam heating of the hot gas after allowance for evaporation,
cooling etc. That is, does the energetics of the beam input implied by
real HXR data tally with the heating (or cooling) of the gas as inferred
from real SXR data.

\begin{figure}[tb]
\centering
\includegraphics[height=6.5cm]{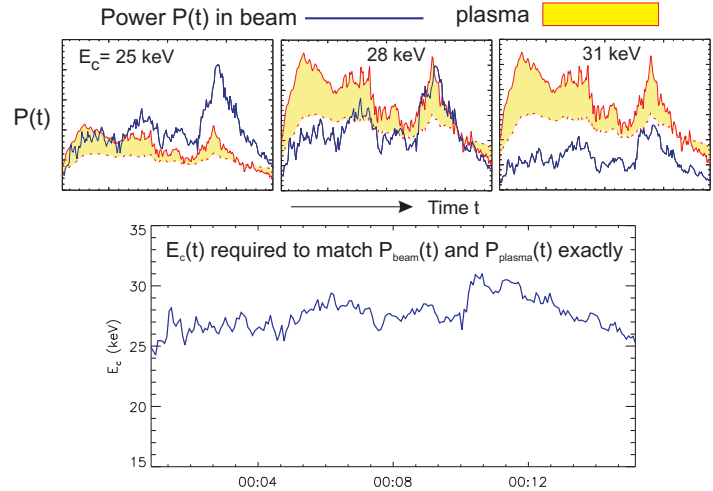}
\caption{Empirical Neupert effect in SXR and HXR light curves for
the M3.2 flare of 2002 April 14/15 translated into SXR plasma power-in and
power-requirement for a single loop, according to \cite{ver:al-05}. Top panels: 
Comparison of the actual power in the hot plasma $P_{\rm plasma}$ required 
to explain the observed SXR flux (minimum-maximum estimate: shaded area) and 
the electron beam power $P_{\rm beam}$ (solid line) calculated for different 
values of the low cutoff energy $E_c$. No value for $E_c$ yields a good match 
between the $P_{\rm plasma}(t)$ and $P_{\rm beam}(t)$ curves. Bottom panel: 
If we allow the low cutoff energy $E_c$ to change during a flare and see how 
it has to vary in order that $P_{\rm plasma}(t)$ and $P_{\rm beam}(t)$ derived 
from observations exactly match at each time step, then it is found that only 
small changes in $E_c$ are necessary.
}
\label{neupertfig}
\end{figure}

Using a rather crude model of energy transport, but one which
should correct the light curves in the right sense at least, Veronig et
al.\ \cite{ver:al-05} studied this for several flares and found the
surprising result that the power into the SXR plasma is less well
correlated with the beam power than are the raw SXR and HXR light
curves -- Figure~\ref{neupertfig}.  They then discuss reasons for this 
result including the possibility that models involving beam-heating of a
single monolithic loop structure may be invalid. In line
with the findings of Fletcher and Hudson \cite{fle:hud-02} mentioned
in Section~\ref{image}, multiple small loop events might offer an
alternative and more realistic description of the global flare emission.
These, however,  also face serious problems since the instantaneous
electron rate (s$^{-1}$) is fixed by the HXR burst intensity and if one
decreases the instantaneous area of injection to that of a small
subloop elements one quickly reaches the point where there are simply
not enough electrons for imaginable densities. In fact, filamentary
beams cannot be less than around 0.1 of the loop radius thick for this
reason, as well as such beams and/or their return currents being
unstable. This is far thicker than the scale of current sheets.

\section{Spectral Discoveries and Issues}
\label{spectra}

\begin{figure}[tb]
\centering
\includegraphics[height=9.3cm]{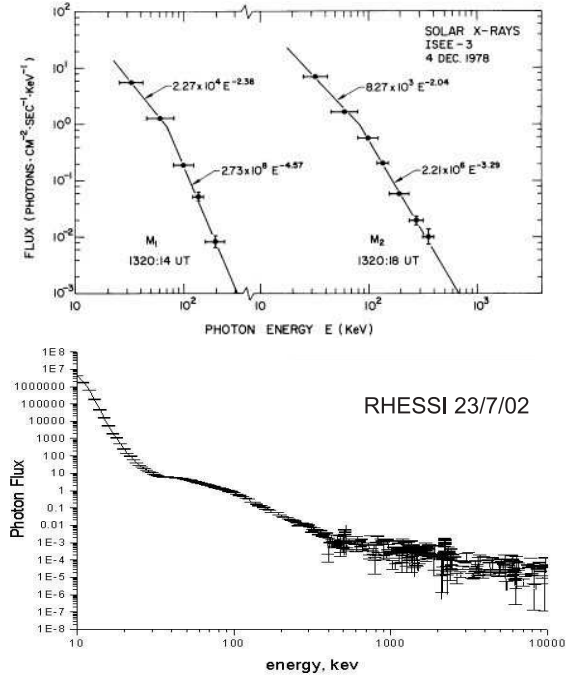}
\caption{Comparison of HXR spectral resolution pre- and post-RHESSI.The
improvement between the 1980s (ISEE etc.) and now (RHESSI) in spectral
energy points/bins from $\Delta \epsilon/\epsilon \sim 0.3$ to
$\sim 0.1$--0.001 over the 10--1000 keV range is huge and has enabled
the first systematic objective inference of electron spectra from their
HXR bremsstrahlung spectra since it was first proposed by Brown
\cite{bro-71}.
}
\label{spectrafig}
\end{figure}

Figure~\ref{spectrafig} shows dramatically the sea-change which
RHESSI's Ge detectors have brought to flare HXR spectrometry, namely an 
increase in resolution from tens of keV to around 1~keV, enabling detailed 
spectral analysis of the bremsstrahlung continuum and resolution of 
individual $\gamma$-ray lines.  The importance of this, emphasised for decades
(e.g. \cite{cra:bro-86}) lies in the fact that the mean source electron
spectrum is essentially the derivative of the photon spectrum
(deconvolved through the bremsstrahlung cross-section) and that the
injected electron spectrum is the further deconvolution 
($\sim$differentiation) of the mean source electron spectrum through 
particle transport smearing effects.

To see this we summarise the derivation of these relationships here.

In the approximation of isotropic emission, the hard X-ray photon
spectrum $I(\epsilon)$ in solar flares is a
convolution of the (density weighted volumetric) {\it mean source
electron spectrum} ${\overline F}(E)$ and the cross-section
$Q(\epsilon,E)$ for production of a photon of energy $\epsilon$ by
an electron of energy $E$, viz.
\begin{equation}
I(\epsilon) = C \, \int_\epsilon^\infty {\overline F}(E)
Q(\epsilon,E) \, dE, \label{defFbar}
\end{equation}
where $C$ is a constant \cite{bro-71,bro:al-03}.

Thus to find $\overline F(E)$ from $I(\epsilon)$ we have to
solve/invert this integral equation which is always rather
unstable to noise in $I(\epsilon)$. Put another way, the integral
involved smears out features of $\overline F(E)$ in emitting the
observable $I(\epsilon)$. A clear example is in the approximation
(Kramers) $Q\sim1/(E\epsilon)$ which leads to the explicit
derivative solution
\begin{equation}
   \overline F(E)\sim
%-E\\left[frac{d(I\epsilon)}d{\epsilon}\right]_{\epsilon=E}
-E\left[\frac{d(I\epsilon)}{d\epsilon}\right]_{\epsilon=E}
\end{equation}
and differentiating data always magnifies high frequency noise
\cite{kon:mac-05}. %Kontar and MacKinnon, 2005).

\begin{figure}[tb]
\centering
\includegraphics[height=5.3cm]{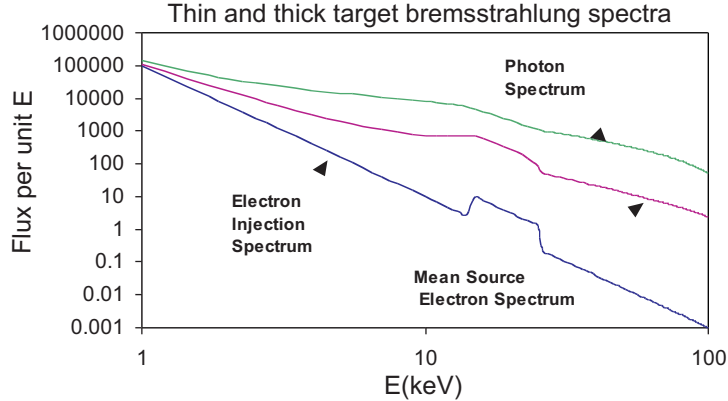}
\caption{
Two stage smearing of a `top-hat' feature on a power-law injected
electron spectrum $F_{0}(E_{0})$ via collisional thick target transport
to give mean source (`thin target') spectrum $\bar{F}(E)$ and via
bremsstrahlung cross section to give the photon spectrum $I(\epsilon)$.
}
\label{tophat}
\end{figure}

In turn, $\overline F(E)$ is related to the electron injection
rate spectrum  $F_0(E_0)$ through the properties of electron
propagation in the source. In the case of purely collisional
transport the relation is \cite{bro:ems-88}
%(Brown and Emslie 1986) 1988??

\begin{equation}
   \overline F(E) \sim E \int_E^\infty F_0(E_0) dE_0
\end{equation}
(so that for a pure power law $\overline F(E)$ is two powers harder
than $F_0(E_0)$).
Consequently details in $F_0(E_0)$ are further smeared out in
$ \overline F(E) $ and the solution for $F_0(E_0)$ for given
$\overline F(E)$ is

\begin{equation}
   F_0(E_0) \sim -\left[\frac{d(\overline F(E)/E)}{dE}\right]_{E=E_0}
\end{equation}
so that $F_0(E_0)$ is sensitive to noise in $ \overline F(E) $ and
extremely sensitive to noise in $I(\epsilon)$
This is illustrated in Figure~\ref{tophat} where a
`top hat' feature in the injected electron spectrum is seen to be
smeared in the mean source spectrum and smeared further in the final
observed bremsstrahlung photon spectrum.

\begin{figure}[tb]
\centering
\includegraphics[height=6cm]{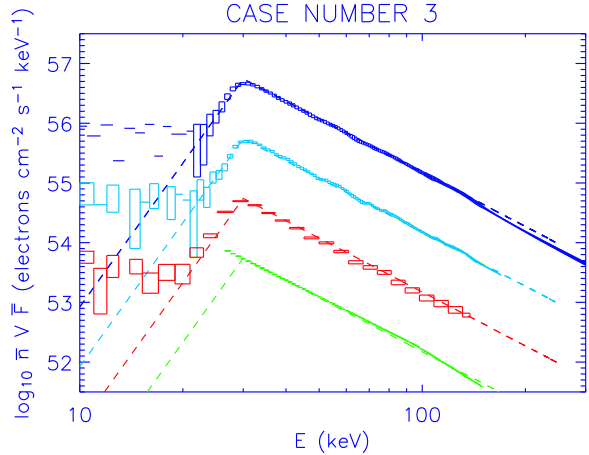}
\caption{One early result of blind tests of spectral
reconstruction algorithms for deriving mean source electron
spectra from noisy photon spectra by regularisation/smoothing
techniques. The four curves, displaced for clarity, are the
results of four different algorithms \cite{joh:lin-92, pia:al-03,
kon:al-04, hol-03} while the dashed curves are the input function
to be recovered. The target electron spectrum was unknown to the
data `inverters' till all results were in, so as to test the
objectivity and consistency of the methods. \label{invertrial} }
\end{figure}

Working backward from the photon data thus requires the careful use
of regularisation algorithms to suppress the effects of spurious
data noise amplification \cite{cra:bro-86} and a great deal of
effort has gone into perfecting such methods \cite{tho:al-92,
joh:lin-92, pia-94, pia:al-03, mas:al-03}. Recent work has aimed at
testing the reliability of these methods \cite{kon:al-04,
kon:al-05a} by applying them `blind' to hypothetical photon spectra
generated by models of electron spectra unknown to the data
analysers.  Overall all the methods prove to be highly reliable, as
can be seen from the example in Figure~\ref{invertrial} the only
problematic regimes being where the source electron spectrum has low
electron numbers in some energy range.  In such regimes the photon
spectrum at energy $E$ is dominated by emission from electrons of
energy $E \gg \epsilon$, rather than  $E \simeq \epsilon$, and so is
a rather poor diagnostic of electron fluxes at such energies. In
practice photon spectra and hence electron spectra are, on the
whole, quite steeply decreasing at all energies so that the electron
spectra should be quite well recovered at most $E$ for most events.

\begin{figure}[tb]
\centering
\includegraphics[height=6.5cm]{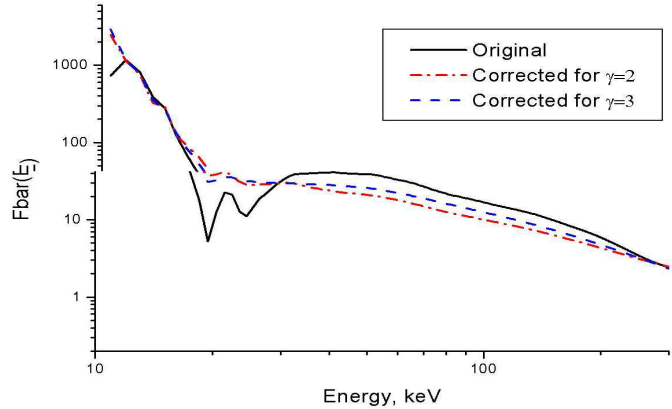}
\caption{Possible removal of inferred `dip' in electron spectrum
$\bar{F}(E)$ by albedo correction of photon spectrum \cite{kon:al-05b}.
When the observed photon spectrum is assumed to be solely due to primary
bremsstrahlung emission the inferred electron spectrum $\overline  F(E)$
for some RHESSI flares showed a dip around 30 keV. But when a correction
was applied to remove the contribution from photospherically
backsacattered
downgoing photons, the dip vanished.
}
\label{edalbedo}
\end{figure}

In a number of events, however, and most notably in the event of
July 23, 2002, the observed HXR spectrum shows locally hard/small
spectral index regions in the 30--60 keV range. When deconvolved,
such photon spectra yield mean source electron spectra with a
non-monotonic `dips' of a kind hitherto totally unknown and of
potentially great importance since their presence might rule out
the canonical thick target model with purely collisional transport
which can only produce a photon spectrum of local spectral index
$\gamma \ge -1$ \cite{bro:ems-88}. It is therefore of vital
importance to test the reality of such features as originating in
the flare electrons themselves rather than in some secondary
process. To date, instrumental origins such as pulse pile up have
not been entirely ruled out while a major issue is the
contribution to the photon spectrum of photospheric back scatter
\cite{ale:bro-02} which is important in the 30--60 keV range. This
arises from the Compton scattering and photoelectric absorption of
downward directed photons from both free and bound electrons.
Preliminary analysis of this effect by Kontar and others suggests
that it can result in inference of a spurious electron spectral
dip in the 30--40 keV range -- Figure~\ref{edalbedo} -- but that
the July 23, 2002 feature $\sim$50--60 keV  might be too high in
energy to be attributable to albedo.

\begin{figure}[tb]
\centering
\includegraphics[height=6cm]{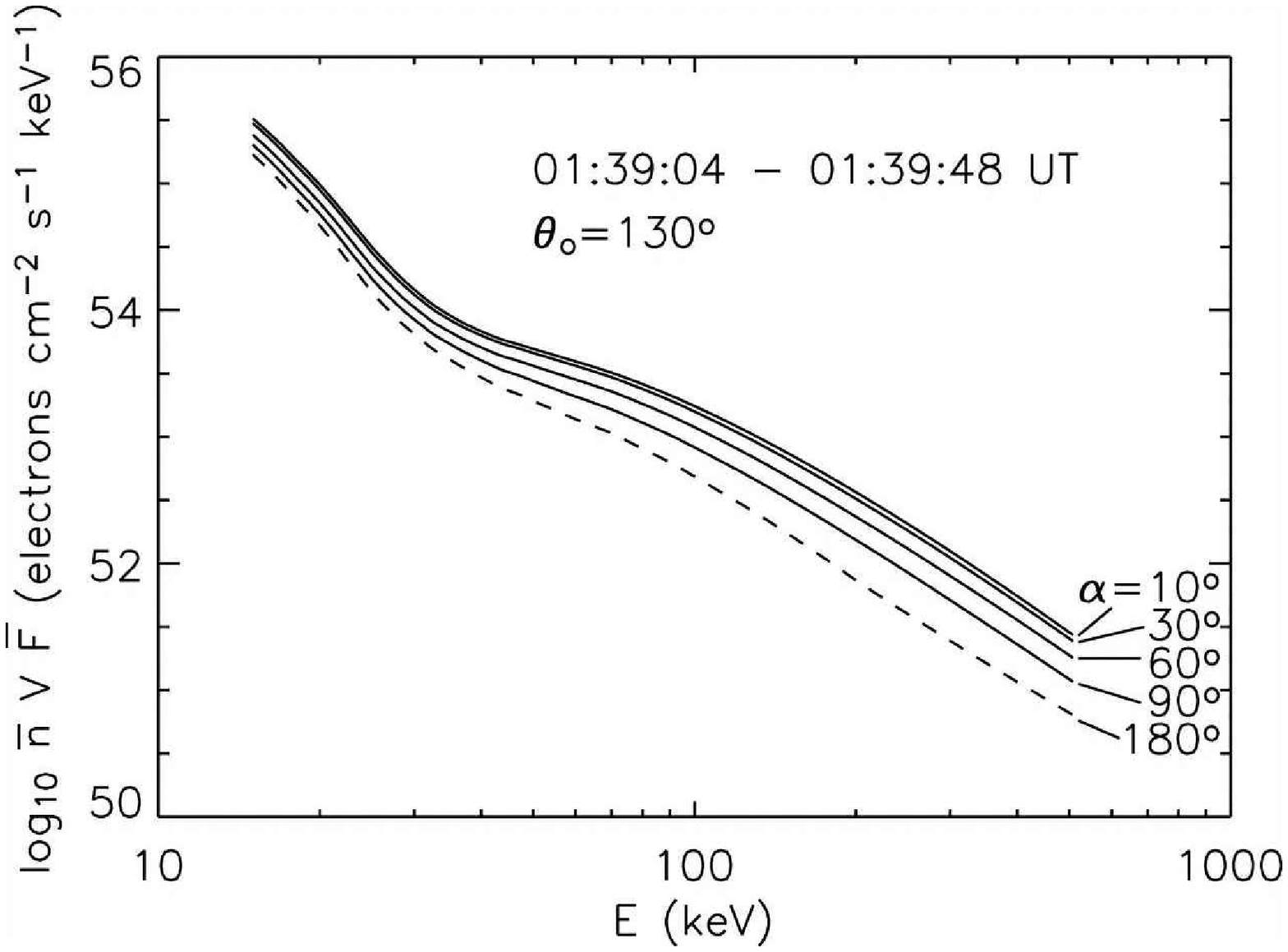}
\caption{Effect of bremsstrahlung cross-section anisotropy on the
inferred $\bar{F}(E)$ for various degrees of source electron
anisotropy, from \cite{mas:al-04}. What this shows is how the inferred source 
electron spectrum $ \overline F(E) $ varies for a variety of assumptions on the
anisotropy of the electron using the full anisotropic bremsstrahlung
cross-section. Note that, as well as the shape of $ \overline F(E) $ there 
is a large variation in the absolute value and hence in the total 
number and energy of the electrons. These results are for theoretical 
primary sources and are not confused by albedo effects.}
\label{massone}
\end{figure}

The importance of albedo in the observed signal also depends on
the extent of the directivity (downward beaming) of the primary
HXR source \cite{kan:al-80}. Even in the absence of albedo,
source directivity affects the inferred slope and flux of the
source electrons -- Massone et al. \cite{mas:al-04} --
Figure~\ref{massone} -- thus much work remains to be done before
we can be confident of inferred electron spectra.

A wholly different but equally intriguing RHESSI `discovery',
demanding rethinking of ideas, is the relationship of the mean
spectral slope of HXR spectra from HXR source electrons at the sun
to the slope of interplanetary electrons at the earth. RHESSI and WIND
have allowed Krucker and Kontar (private communication) to confirm,
with much improved data, earlier results of Lin et al. \cite{lin:al-82} and
Lin \cite{lin-85, lin-93}.

If all electrons are accelerated in the corona (as opposed to say
separate acceleration sites in the corona and chromosphere) with
spectral index $\delta$ then one would expect that: (a) upward interplanetary
electrons would arrive at the earth with (scatter-free) index $\delta$; (b) HXR's
emitted from downward injection of such electrons into a dense collisional
thick target \cite{bro-71} would produce HXRs of index $\delta - 1$;
(c)~HXR's emitted around the tenuous acceleration site would have
thin target index $\delta + 1$. Krucker and Kontar, like Lin earlier,
found that real flares are much closer to regime~(c) than to regime~(b).

This seeming violation of the predictions of the basic collisional
thick target model is contrary to the RHESSI imaging results cited in
Section~\ref{image}  which seems in broad support of that
thick target model -- i.e. we do see footpoints. It is possible that
interplanetary electron propagation is not scatter-free or that thick
target electrons undergo non-collisional losses but in either interpretation 
it is strange that the transport effects are just such as to yield data
close to thin target situation (c). Time indeed for some rethinking!

\section{CONCLUSIONS}
\label{conclude}

RHESSI data constitute the greatest breakthrough in flare fast
particle studies since the first HXR detectors were launched over
30 years ago. The results will pose `rethink' challenges for an
entire new generation of solar physicists, all the more so when
considered in the wider context of multi-wavelength data,
especially in the complementary radio regime to which CESRA is
dedicated.

\acknowledgement{This work was supported by a PPARC Rolling Grant
(JCB, EPK) and a Visitor Grant from the Royal Society of Edinburgh
(AMV). We are grateful to Alec MacKinnon for help with the
manuscript.}

\bibliographystyle{cl2emult}

\bibliography{cesra_jcbrown}

\begin{thebibliography}{50.}
\addcontentsline{toc}{chapter}{References}
\newcommand{\enquote}[1]{``#1''}
\expandafter\ifx\csname url\endcsname\relax
  \def\url#1{{\tt #1}}\fi
\expandafter\ifx\csname urlprefix\endcsname\relax\def\urlprefix{URL }\fi

\bibitem{lin:al-02}
R.P. {Lin}, B.R. {Dennis}, G.J. {Hurford}, D.M. {Smith}, A.~{Zehnder}, P.R.
  {Harvey}, D.W. {Curtis}, D.~{Pankow}, P.~{Turin}, M.~{Bester},
  A.~{Csillaghy}, M.~{Lewis}, N.~{Madden}, H.F. {van Beek}, M.~{Appleby},
  T.~{Raudorf}, J.~{McTiernan}, R.~{Ramaty}, E.~{Schmahl}, R.~{Schwartz},
  S.~{Krucker}, R.~{Abiad}, T.~{Quinn}, P.~{Berg}, M.~{Hashii}, R.~{Sterling},
  R.~{Jackson}, R.~{Pratt}, R.D. {Campbell}, D.~{Malone}, D.~{Landis}, C.P.
  {Barrington-Leigh}, S.~{Slassi-Sennou}, C.~{Cork}, D.~{Clark}, D.~{Amato},
  L.~{Orwig}, R.~{Boyle}, I.S. {Banks}, K.~{Shirey}, A.K. {Tolbert},
  D.~{Zarro}, F.~{Snow}, K.~{Thomsen}, R.~{Henneck}, A.~{Mchedlishvili},
  P.~{Ming}, M.~{Fivian}, J.~{Jordan}, R.~{Wanner}, J.~{Crubb}, J.~{Preble},
  M.~{Matranga}, A.~{Benz}, H.~{Hudson}, R.C. {Canfield}, G.D. {Holman},
  C.~{Crannell}, T.~{Kosugi}, A.G. {Emslie}, N.~{Vilmer}, J.C. {Brown},
  C.~{Johns-Krull}, M.~{Aschwanden}, T.~{Metcalf}, A.~{Conway}: {Solar Phys.}
  {\bf 210\/}, 3--32 (2002)

\bibitem{den:al-05}
B.R. {Dennis}: \enquote{{Review of selected RHESSI solar results}}, in  {\em
  The High Energy Solar Corona: Waves, Eruptions, Particles\/} (this volume)

\bibitem{bro-71}
J.C. {Brown}: {Solar Phys.} {\bf 18\/}, 489 (1971)

\bibitem{mas:al-94}
S.~{Masuda}, T.~{Kosugi}, H.~{Hara}, S.~{Tsuneta}, Y.~{Ogawara}: {Nature} {\bf
  371\/}, 495 (1994)

\bibitem{kan-83}
S.R. {Kane}: {Solar Phys.} {\bf 86\/}, 355--365 (1983)

\bibitem{lin:sch-87}
R.P. {Lin}, R.A. {Schwartz}: {Astrophys. J.} {\bf 312\/}, 462--474 (1987)

\bibitem{hur:al-03}
G.J. {Hurford}, R.A. {Schwartz}, S.~{Krucker}, R.P. {Lin}, D.M. {Smith},
  N.~{Vilmer}: {Astrophys. J. Letts.} {\bf 595\/}, L77--L80 (2003)

\bibitem{ems:al-04}
A.G. {Emslie}, J.A. {Miller}, J.C. {Brown}: {Astrophys. J. Letts.} {\bf 602\/},
  L69--L72 (2004)

\bibitem{bro:mcc-76}
J.C. {Brown}, A.N. {McClymont}: Solar Phys. {\bf 49\/}, 329--342 (1976)

\bibitem{mac:al-85}
A.L. {MacKinnon}, J.C. {Brown}, J.~{Hayward}: Solar Phys. {\bf 99\/}, 231--262
  (1985)

\bibitem{bro:al-02}
J.C. {Brown}, M.J. {Aschwanden}, E.P. {Kontar}: {Solar Phys.} {\bf 210\/},
  373--381 (2002)

\bibitem{asc:al-02}
M.J. {Aschwanden}, J.C. {Brown}, E.P. {Kontar}: {Solar Phys.} {\bf 210\/},
  383--405 (2002)

\bibitem{ems:al-03}
A.G. {Emslie}, E.P. {Kontar}, S.~{Krucker}, R.P. {Lin}: {Astrophys. J. Letts.}
  {\bf 595\/}, L107--L110 (2003)

\bibitem{fle:hud-02}
L.~{Fletcher}, H.S. {Hudson}: {Solar Phys.} {\bf 210\/}, 317--321 (2002)

\bibitem{ver:bro-04}
A.M. {Veronig}, J.C. {Brown}: {Astrophys. J. Letts.} {\bf 603\/}, L117--L120
  (2004)

\bibitem{sui:al-04}
L.~{Sui}, G.D. {Holman}, B.R. {Dennis}: {Astrophys. J.} {\bf 612\/}, 546--556
  (2004)

\bibitem{kos:al-94}
T.~{Kosugi}, T.~{Sakao}, S.~{Masuda}, H.~{Hara}, T.~{Shimizu}, H.S. {Hudson}:
  \enquote{{Hard and Soft X-ray Observations of a Super-Hot Thermal Flare of 6
  February, 1992}}, in  {\em Proceedings of Kofu Symposium\/} (1994), pp.
  127--129

\bibitem{kan:hur-03}
S.R. {Kane}, G.J. {Hurford}: Advances in Space Research {\bf 32\/}, 2489--2493
  (2003)

\bibitem{kan:al-04}
S.R. {Kane}, J.M. {McTiernan}, K.~{Hurley}: American Astronomical Society
  Meeting Abstracts {\bf 204\/} (2004)

\bibitem{ram:al-97}
R.~{Ramaty}, N.~{Mandzhavidze}, C.~{Barat}, G.~{Trottet}: {Astrophys. J.} {\bf
  479\/}, 458 (1997)

\bibitem{bro:al-00}
J.C. {Brown}, S.~{Krucker}, M.~{G{\" u}del}, A.O. {Benz}: {Astron. Astrophys.}
  {\bf 359\/}, 1185--1194 (2000)

\bibitem{kru:al-02}
S.~{Krucker}, S.~{Christe}, R.P. {Lin}, G.J. {Hurford}, R.A. {Schwartz}: {Solar
  Phys.} {\bf 210\/}, 445--456 (2002)

\bibitem{ver:al-05}
A.M. {Veronig}, J.C. {Brown}, B.R. {Dennis}, R.A. {Schwartz}, L.~{Sui}, A.K.
  {Tolbert}: Astrophys. J. {\bf 621\/}, 482--497 (2005)

\bibitem{neu-68}
W.M. {Neupert}: Astrophys. J. Letts. {\bf 153\/}, L59--64 (1968)

\bibitem{den:zar-93}
B.R. {Dennis}, D.M. {Zarro}: Solar Phys. {\bf 146\/}, 177--190 (1993)

\bibitem{lee:al-95}
T.T. {Lee}, V.~{Petrosian}, J.M. {McTiernan}: Astrophys. J. {\bf 448\/},
  915--924 (1995)

\bibitem{mct:al-99}
J.M. {McTiernan}, G.H. {Fisher}, P.~{Li}: Astrophys. J. {\bf 514\/}, 472--483
  (1999)

\bibitem{ver:al-02}
A.~{Veronig}, B.~{Vr{\v s}nak}, B.R. {Dennis}, M.~{Temmer}, A.~{Hanslmeier},
  J.~{Magdaleni{\' c}}: Astron. Astrophys. {\bf 392\/}, 699--712 (2002)

\bibitem{li:al-93}
P.~{Li}, A.G. {Emslie}, J.T. {Mariska}: Astrophys. J. {\bf 417\/}, 313--319
  (1993)

\bibitem{li:al-97}
P.~{Li}, J.M. {McTiernan}, A.G. {Emslie}: Astrophys. J. {\bf 491\/}, 395 (1997)

\bibitem{cra:bro-86}
I.J.D. {Craig}, J.C. {Brown}: {\em {Inverse Problems in Astronomy}\/} (Adam
  Hilger, 1986)

\bibitem{bro:al-03}
J.C. {Brown}, A.G. {Emslie}, E.P. {Kontar}: Astrophys. J. Letts. {\bf 595\/},
  L115--L117 (2003)

\bibitem{kon:mac-05}
E.P. {Kontar}, A.L. {MacKinnon}: Solar Phys. {\bf 227\/}, 299--310 (2005)

\bibitem{bro:ems-88}
J.C. {Brown}, A.G. {Emslie}: Astrophys. J. {\bf 331\/}, 554--564 (1988)

\bibitem{tho:al-92}
A.M. {Thompson}, J.C. {Brown}, I.J.D. {Craig}, C.~{Fulber}: {Astron.
  Astrophys.} {\bf 265\/}, 278--288 (1992)

\bibitem{joh:lin-92}
C.M. {Johns}, R.P. {Lin}: Solar Phys. {\bf 137\/}, 121--140 (1992)

\bibitem{pia-94}
M.~{Piana}: {Astron. Astrophys.} {\bf 288\/}, 949--959 (1994)

\bibitem{pia:al-03}
M.~{Piana}, A.M. {Massone}, E.P. {Kontar}, A.G. {Emslie}, J.C. {Brown}, R.A.
  {Schwartz}: {Astrophys. J. Letts.} {\bf 595\/}, L127--L130 (2003)

\bibitem{mas:al-03}
A.M. {Massone}, M.~{Piana}, A.~{Conway}, B.~{Eves}: {Astron. Astrophys.} {\bf
  405\/}, 325--330 (2003)

\bibitem{kon:al-04}
E.P. {Kontar}, M.~{Piana}, A.M. {Massone}, A.G. {Emslie}, J.C. {Brown}: Solar
  Phys. {\bf 225\/}, 293--309 (2004)

\bibitem{kon:al-05a}
E.P. {Kontar}, A.G. {Emslie}, M.~{Piana}, A.M. {Massone}, J.C. {Brown}: Solar
  Phys. {\bf 226\/}, 317--325 (2005)

\bibitem{hol-03}
G.D. {Holman}: Astrophys. J. {\bf 586\/}, 606--616 (2003)

\bibitem{ale:bro-02}
R.C. {Alexander}, J.C. {Brown}: {Solar Phys.} {\bf 210\/}, 407--418 (2002)

\bibitem{kon:al-05b}
E.P. {Kontar}, A.~{MacKinnon}, J.C. {Brown}: Astron. Astrophys. {\bf
  submitted\/} (2005)

\bibitem{kan:al-80}
S.R. {Kane}, K.A. {Anderson}, W.D. {Evans}, R.W. {Klebesadel}, J.G. {Laros}:
  {Astrophys. J.} {\bf 239\/}, L85--L88 (1980)

\bibitem{mas:al-04}
A.M. {Massone}, A.G. {Emslie}, E.P. {Kontar}, M.~{Piana}, M.~{Prato}, J.C.
  {Brown}: {Astrophys. J.} {\bf 613\/}, 1233--1240 (2004)

\bibitem{lin:al-82}
R.P. {Lin}, R.A. {Mewaldt}, M.A.I. {van Hollebeke}: Astrophys. J. {\bf 253\/},
  949--962 (1982)

\bibitem{lin-85}
R.P. {Lin}: Solar Phys. {\bf 100\/}, 537--561 (1985)

\bibitem{lin-93}
R.P. {Lin}: Advances in Space Research {\bf 13\/}, 265--273 (1993)

\end{thebibliography}

%\begin{thebibliography}

\end{document}